\documentclass[fleqn,10pt,twocolumn]{wlscirep}
\usepackage[utf8]{inputenc}
\usepackage[T1]{fontenc}
\usepackage{newunicodechar}
\newunicodechar{π}{\pi}
\newunicodechar{−}{-}
\newunicodechar{∼}{\sim}
\newunicodechar{σ}{\sigma}
\usepackage{array}
\usepackage{float}
\usepackage[version=4]{mhchem}
\DeclareCaptionLabelSeparator{none}{ }
\usepackage[font=small,labelfont=bf,
   justification=justified,
   format=plain]{caption} 
\usepackage[figurename=Fig.]{caption}

\title{Prediction of an alternative route to polymeric carbon dioxide: A metastable energetic material}

\author[1]{Reetam Paul}
\author[1]{Jonathan C. Crowhurst}
\author[1,*]{Stanimir A. Bonev}
\affil[1]{Lawrence Livermore National Laboratory, 7000 East Avenue, Livermore, CA 94550}
\affil[*]{bonev2@llnl.gov}


\begin{abstract}
The use of pressure to obtain new materials that can be recovered under ambient conditions is a central problem in high-pressure physics.  Despite decades of research, this goal has only been achieved in the laboratory for a few notable examples, such as diamond and cubic boron nitride. An area of significant interest is the transformation under compression of light-element molecular compounds to extended covalent-bonded (polymeric) solids. Among them, CO\textsubscript{2} has been extensively studied because of its status as a prototypical simple molecular system with a rich phase diagram and due to its fundamental role in Earth's physics and chemistry. One of its polymeric crystalline phases, accessible at extreme pressures and temperatures, has been recently quenched to ambient pressure, but below room temperature. Here we report \textit{ab initio} calculations predicting that isothermal compression of a carbon monoxide and oxygen mixture (CO+O\textsubscript{2}), rather than the compound CO\textsubscript{2}, lowers the onset of C-polymerization at room temperature from $\sim$118~GPa to $\sim$7~GPa (complete by $\sim$23 GPa). Moreover, it leads to the formation of an intrinsically different polymer with enhanced metastability. We predict that this dense phase is an energetic material which can potentially be recovered to ambient pressure and temperature.
\end{abstract}
\begin{document}

\flushbottom
\maketitle

\thispagestyle{empty}


In the high-pressure sciences, there have been discoveries of exotic crystalline phases of even simple elements such as Ca \cite{novoselov2020weak}, Na \cite{ma2009transparent}, and Li \cite{neaton1999pairing}. However, almost without fail, thermodynamic or kinetic effects lead to the breakdown of such exotic phases upon decompression (relaxation) and their practical applications as superhard \cite{hilleke2022materials} or high-energy-density (HED) materials have been precluded, albeit with a few exceptions such as diamond~\cite{sundqvist2021carbon} and cubic boron nitride \cite{datchi2007equation,goncharov2007thermal}. 

The confluence of high-pressure physics and energetic materials research has led to the prediction that the high-pressure ($\sim$110 GPa) and high-temperature ($\sim$2000~K) synthesis of cubic gauche nitrogen (\textit{cg}-N, space group \textit{I}2\textsubscript{1}3) \cite{eremets2004single,peiris2008static} would lead to an energetic material having upper end estimates of energy content more than twice that of TNT ($>$10 kJ/g), undoubtedly the most energetic non-nuclear composition in existence. Practical use would, of course, depend on developing a method for synthesizing the material at ambient pressure and temperature, which has very recently been suggested \cite{xu2024free}, as all attempts at decompression-based recovery have failed. The materials further explored in this category include compressed phases of N-rich compounds~\cite{bini2000high,eremets2004polymerization,mattson2004prediction,eremets2004single,gregoryanz2007high,pickard2009high}, CO\textsubscript{2}~\cite{yoo1999crystal,bonev2003high,giordano2006melting,santoro2006amorphous,giordano2007molecular,iota2007six,sun2009high,datchi2009structure,sun2009high,giordano2010equation,santoro2012partially,datchi2014structure,datchi2017polymeric,dziubek2018crystalline}, CH\textsubscript{4 } \cite{gao2010dissociation}, NH\textsubscript{3}~\cite{ninet2008high,ojwang2012melting}, and CO~\cite{lipp2005high,bonev2021energetics,scelta2023high}. They form extended covalent-bonded (polymeric) solids that, if recovered to ambient conditions, would be in a high-energy metastable state. A subsequent transition to a thermodynamically stable phase would release a large amount of energy. Naturally, much research has focused on attempts to make the synthesis conditions more accessible while increasing the HED content and ambient-condition metastability of the high-pressure phases. These requirements are often conflicting. 

Carbon dioxide was one of the first molecular systems that was shown to polymerize under compression~\cite{iota1999quartzlike,datchi2012structure}. Given its fundamental nature and importance for understanding the physics and chemistry of the Earth, its phase diagram has been of particular interest to the high-pressure research community~\cite{iota2001phase,yoo2011phase,cogollo2020ab,bonev2003high}. Below 200~K and at ambient pressure, CO\textsubscript{2} exists in phase I as a molecular solid ‘dry ice’ cubic structure (\textit{Pa}3). At pressures above $\sim$12~GPa, CO\textsubscript{2}-I transforms to the molecular CO\textsubscript{2}-III (\textit{Cmca}), CO\textsubscript{2}-II (\textit{P}4\textsubscript{2}/\textit{mnm}) or CO\textsubscript{2}-IV (\textit{Pbcn}), in increasing order of temperature~\cite{yoo1999crystal,bonev2003high}. Thereafter, the picture becomes somewhat uncertain. A fully covalent-bonded CO\textsubscript{2}-V structure was obtained above 40-60 GPa with heating up to 1800~K, and reported as either tridymite (high temperature polymorph of silica/SiO\textsubscript{2}, tetrahedrally bonded/fourfold-coordinated)-type (\textit{P}2\textsubscript{1}2\textsubscript{1}2\textsubscript{1}) at higher temperatures \cite{holm2000theoretical} or an extended-solid cristobalite-type (\textit{I}4-2\textit{d}) at lower temperatures \cite{kim2018transformation,santoro2012partially}. Metadynamics simulations \cite{sun2009high} starting from CO\textsubscript{2}-III at $\sim$80 GPa and $<$300~K suggest an intermediate \textit{Pbca}  ultimately evolving to an $\alpha$-cristobalite-like (\textit{P}4\textsubscript{1}2\textsubscript{1}2) fourfold-coordinated structure, whereas CO\textsubscript{2}-II at $\sim$60 GPa and $\sim$600~K transforms to a fully tetrahedral, layered structure ({\textit{P}$\overline{4}$\textit{m}2). However, experimental observations \cite{iota2007six} reported a fully covalent extended-solid sixfold-coordinated stishovite-type CO\textsubscript{2}-VI (\textit{P}4\textsubscript{2}/\textit{mnm}) obtained directly by compressing CO\textsubscript{2}-II above 50 GPa and at 530-650 K.  There have also been predictions by simulations~\cite{santoro2006amorphous,serra1999pressure} and observations in laser-heated conditions~\cite{tschauner2001new}  above 40 GPa of an amorphous phase called carbonia/\textit{a}-CO\textsubscript{2}.  A combination of experiments and simulations~\cite{montoya2008mixed} have shown that the actual nature of CO\textsubscript{2} in the 40-100 GPa region is a mixture of three- and four-fold coordinated metastable phases, manifesting as \textit{a}-CO\textsubscript{2}, rather than any particular well-defined fully tetrahedral phase.  Recent attempts to recover polymeric crystalline CO\textsubscript{2}-V ($P2_12_12_1$) \cite{yong2016crystal} have shown that the recovered product decays to dry ice/CO\textsubscript{2}-I (\textit{Pa}$\overline{3}$) at 185 K. As such, recovering polymeric CO\textsubscript{2} to atmospheric pressure at room temperature has been hitherto elusive \cite{datchi2017polymeric}.

In general, pressure-induced polymerization of molecular compounds can be divided into two types, thermodynamic and kinetic transitions. The former takes place between phases that are in thermodynamic equilibrium. The compression of CO\textsubscript{2} and N$_2$ until polymerization falls into this category. Here, the amount of energy stored in the polymeric phases creates a trade-off with the accessibility of synthesis conditions. In the second type, the transition is from or to a phase which is metastable (i.e. a high-energy state). Thus, the kinetic barrier that separates the molecular and polymeric phases can be overcome at relatively low pressure. This is the case for CO, where the polymeric (\textit{p}-CO) rather than the molecular phase is the stable one under ambient conditions, and the onset of the transition is only at around 5~GPa at 300~K. However, the usefulness of \textit{p}-CO as a HED material is limited by the fact that its relevant exothermic reaction, namely the decomposition of \textit{p}-CO to CO$_2^{\mathrm{(\textit{g})}}$ + C, contains a product (C) with a relatively high formation energy \cite{bonev2021energetics}.  

The challenge and a key for discovering a useful material is to combine the benefits of the two types of transitions. Systems such as polymeric CO\textsubscript{2} (\textit{p}-CO\textsubscript{2}) and \textit{cg}-N are compelling due to the significant energy differential between these and their stable molecular phases at ambient pressure. However, to reduce the transition pressures, we wish to arrive at one of these polymeric phases via a kinetic transition from a high-energy state compared to their molecular phases. These considerations have motivated us to investigate the polymerization of a mixture of CO and O\textsubscript{2}, stoichiometrically equivalent to the compound CO\textsubscript{2}.

Another reason for evaluating this system is the expectation that amorphous systems offer more possibilities to achieve high metastability. First, decompression of high-pressure covalent crystalline phases can cause significant bond strains, leading them to become dynamically unstable. Second, kinetic barriers depend on the atomic arrangements, and amorphous solids have higher degrees of freedom for such arrangements.

Here we show that isothermal compression of a mixture of CO and O\textsubscript{2} at 300~K leads to a polymer stoichiometrically equivalent to \textit{p}-CO\textsubscript{2}, but accessible at much lower pressure than when compressing molecular CO\textsubscript{2} (\textit{m}-CO\textsubscript{2}). Importantly, the resulting amorphous polymer is structurally different from the known \textit{p}-CO\textsubscript{2} phases and exhibits higher degree of metastability. Our study indicates that it can be stabilized at even elevated temperatures at near-atmospheric pressure. In the next section we describe the first-principles simulations leading to this prediction, followed by an analysis of structural and thermodynamics properties and metastability, comparative with that of the relevant solid CO\textsubscript{2} phases.

\section*{Results}

\subsection*{Prediction of polymerization and recovery at 300~K}
A gaseous mixture of molecular CO and O\textsubscript{2} was initially equilibrated using \textit{ab initio} molecular dynamics (AIMD) simulations at ambient temperature and pressure conditions. It was then isothermally compressed along the 300~K isotherm. This approach results in the simulation of amorphous, rather than crystalline, polymeric high-pressure phases. It is intended to mimic experimental conditions, and in the case of CO was shown to agree well with measurements~\cite{bonev2021energetics}.

As detailed in the following sections, simulations were performed using (1) different numbers of atoms (as high as 864) to ensure size convergence and (2) various exchange-correlation functionals to confirm that the results are not artifacts of the functional choice. Unless specified otherwise, the discussion refers to the calculations with 432-atom supercells using the PBE exchange-correlation functional with DFT-D3 (zero damping)~\cite{grimme2010consistent} semi-local dispersion forces. A full description of computational details can be found in the Methods section and Tables S1-S2 in Supplementary Discussion 1.

In these simulations, the onset of pressure-induced polymerization is predicted at approximately 7~GPa ($\sim$80\% of ambient cell volume). The transformation is gradual; the early polymeric phase consists of broken chains and transient 4-member C$-$O rings. By $\sim$23 GPa, complete polymerization of the CO molecules is observed (denoted as the \textit{p}\textsubscript{a} phase), characterized by $-$O(CO)$-$ chains linking either planar or non-planar 5-member (C,O) rings and 6-member (C,O) rings. A visualization of the amorphous \textit{p}\textsubscript{a} phase with fully-polymerized C atoms is shown in Fig.~1a. The fraction of carbon and oxygen atoms locked in different bonding configurations in chains and rings is shown in Fig.~1b as a function of the simulation cell size (see Supplementary Discussion 2 for further details). The 432-atom simulations are well-converged with respect to equilibrated pressure and bonding statistics. Finally, upon decompression (simulated for a total of 10.1 ps) from the highest compressed configuration at $\sim$32 GPa along the 300~K isotherm, the \textit{p}\textsubscript{a}  phase is retained all the way to ambient pressure with only a minor a change in the bonding statistics (see Supplementary Figs. S1-S3). Machine-learned molecular dynamics (MLMD) simulations for a 2,926-atom cell with a total simulation time of 2.5~ns also confirm the stability of the amorphous structure (see Supplementary Fig. S6 for details).

To explore the possibility of formation, under isothermal compression, of [CO+O$_2$]-like crystalline structures that are energetically competitive with the \textit{p}\textsubscript{a} phase, we complemented the above approach with constrained metadynamics simulations. In this methodology, the CO and O$_2$ bond lengths were constrained to remain close to their unreacted equilibrium values. The simulations yielded two polymeric crystalline phases: one with an 84-atom unit cell (space group $Pbcn$) and another one with a 36-atom unit cell ($P1$), denoted as \textit{p}\textsubscript{c1} and \textit{p}\textsubscript{c2}, respectively. A visualization of these phases and a comparison of their enthalpies to that of the amorphous \textit{p}\textsubscript{a}  are shown in Fig.~1c. They are energetically favorable to \textit{p}\textsubscript{a} only in a small pressure region below 10~GPa, and when subjected to constant-pressure AIMD, they become disordered (see Supplementary Fig.~4 for visualization). Upon further compression, they amorphize to structures identical to \textit{p}\textsubscript{a}. Therefore, we conclude that while intermediate crystalline phases are possible, isothermal compression at 300~K of a mixture of molecular CO and O\textsubscript{2} leads to the amorphous phase \textit{p}\textsubscript{a} at pressure above 10 GPa, which will be the principal focus of this paper. Note that all these phases are metastable relative to known molecular and polymeric CO$_2$ phases, as will be discussed in the following sections.  

\begin{figure*}[!t]
\centering
\includegraphics[width=0.9\linewidth]{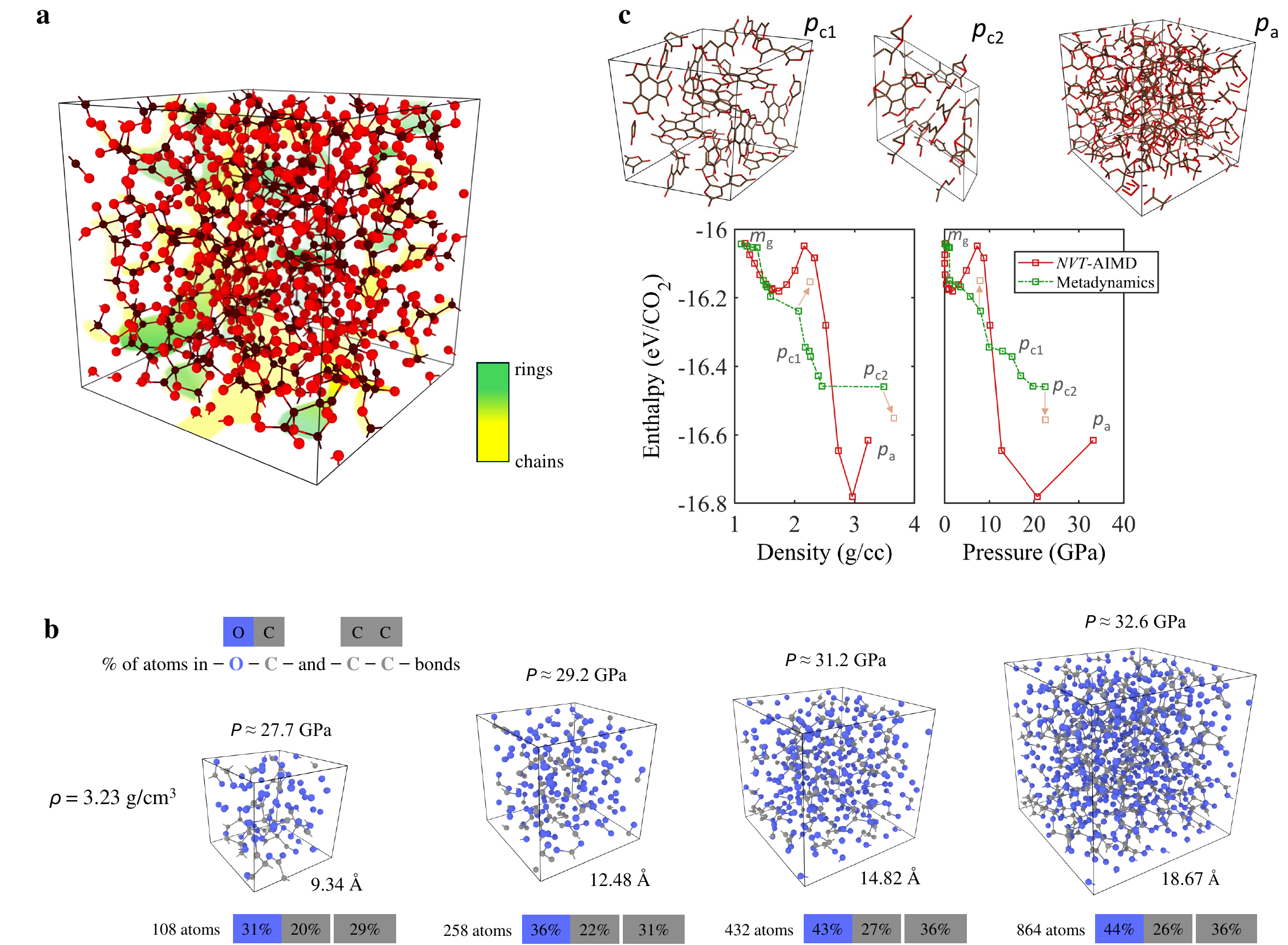}
\begin{center}
\vspace{1cm}
\caption{Results from isothermal compression of CO+O\textsubscript{2} mixtures using \textit{NVT}-AIMD simulations along the \textit{T }= 300~K isotherm. \textbf{a} Polymerized amorphous product in an 864-atom cell at 32.6$\pm$0.7~GPa and $\rho$ = 3.23~g/cc after simulation time of 11 ps at 300~K. Red and brown spheres indicate oxygen and carbon atoms, respectively, with chain and ring regions color-coded as shown.  \textbf{b} Comparison of simulation results obtained with different cell sizes at $\rho$ = 3.23 g/cc. Shown are the percentage of atoms locked in $-$O$-$C$-$ and $-$C$-$C$-$ bonds, and the equilibrated pressure \textit{P}. Convergence is reached with 432-atom simulations. Violet and gray spheres correspond to oxygen and carbon atoms, respectively. The dimensions of the cubic box simulation cells are shown in Angstroms. \textbf{c} Starting from a molecular gaseous mixture of CO and O\textsubscript{2}, labeled as phase \textit{m}\textsubscript{g}, the resulting phases from AIMD and metadynamics searches are shown in perspective view. The polymeric crystalline phases \textit{p}\textsubscript{c1} (84-atom \textit{Pbcn}) and \textit{p}\textsubscript{c2} (36-atom \textit{P}1) were obtained from constrained metadynamics searches. At pressures above 10 GPa, the polymeric amorphous phase \textit{p}\textsubscript{a} is energetically preferred, as can be seen in the enthalpy versus pressure and density plots. The brown arrows and markers point to the change in the enthalpy and density when \textit{NPT}-AIMD simulations are performed on the crystalline phases.}
\end{center}
\vspace{1cm}
\label{fig:main}
\end{figure*}

\begin{figure*}[!t]
\centering
\includegraphics[width=0.75\linewidth]{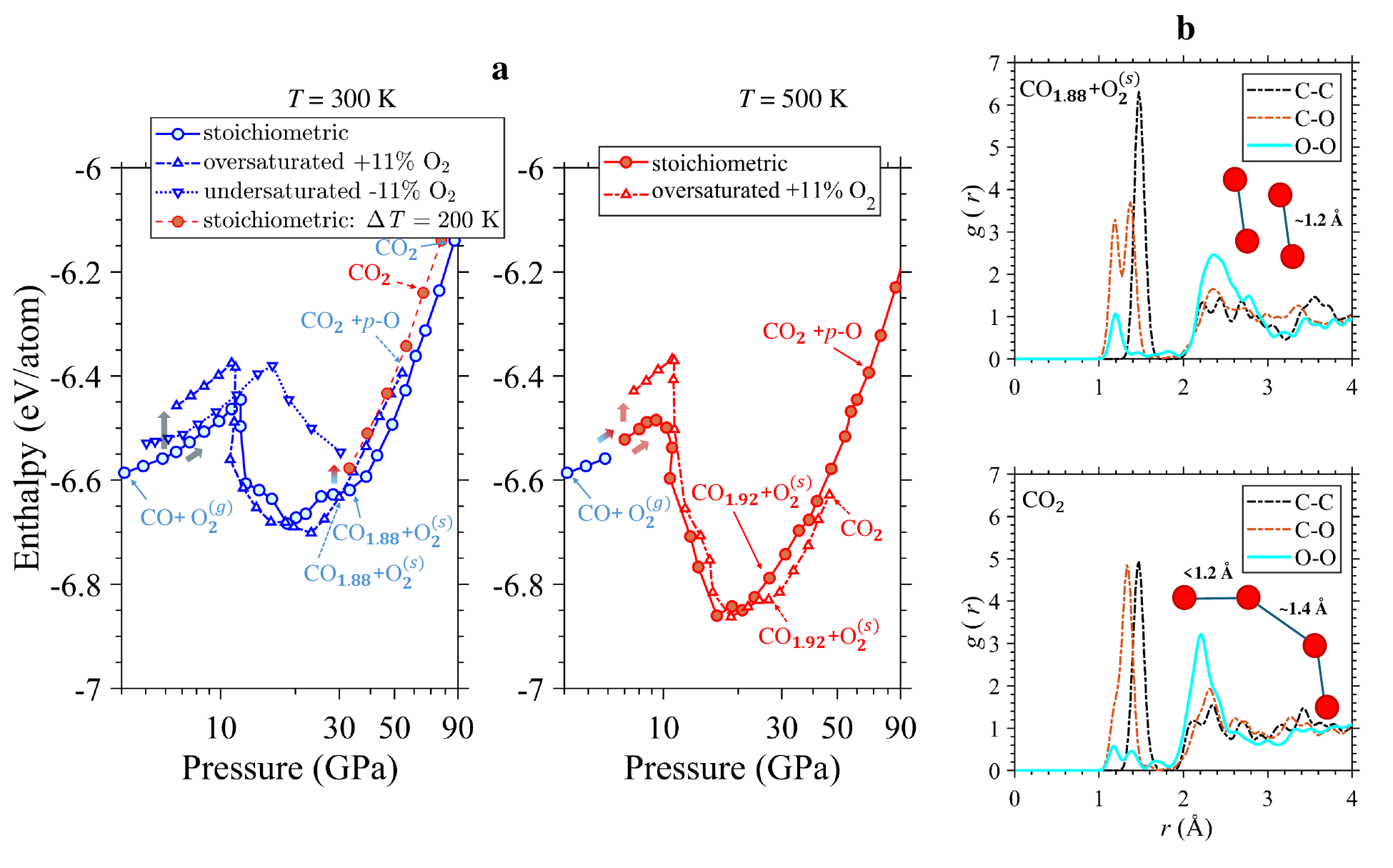}
\begin{center}
\vspace{1cm}
\caption{\textbf{a}  Enthalpy versus pressure plots for 192-atom \textit{NVT}-AIMD simulations along compression trajectories at (left) 300~K and (right) 500~K.
  Plotted are results for systems with stoichiometric CO$_2$ mixtures, as well as such with 11\% over- and under-saturation of oxygen. The red symbols in the left panel indicate a system heated to 500~K at about 32~GPa and then further compressed. The remaining results are for isothermal compressions staring from below 7~GPa.  The labels to selected points indicate the stoichiometry of the polymeric phase achieved at various pressures and the presence of unreacted (O$_2^{\mathrm{(\textit{s})}}$) or polymerized ($p$-O) oxygen.
  \textbf{b} Radial distribution function, \textit{g}(\textit{r}), for CO\textsubscript{1.88}+O$_2^{\mathrm{(\textit{s})}}$ (top) and CO\textsubscript{2} (bottom) along the 300~K compression trajectory at approximately 32 and 88~GPa, respectively. The inset cartoons show how the remaining unreacted O\textsubscript{2} molecules at 32~GPa link up to form chains in fully-polymerized CO+O\textsubscript{2} at 88~GPa.
}
\end{center}
\vspace{1cm}
\label{fig:2phase}
\end{figure*}

A consideration when analyzing the stoichiometry of the compressed  \textit{p}\textsubscript{a} polymer is to determine if there are any residual unreacted molecules remaining in the system. We observe that at around 30~GPa, when the C atoms are fully polymerized, there are a few remaining O$_2$ molecules. Upon further compression, these residual molecules form oxygen chains, and at  $\sim$88 GPa, they attach to the remaining polymer without affecting its structure. We have investigated several different recipes for the formation of polymers with exact CO$_2$ stoichiometry, including over- and undersaturation with O$_2$, compression along a higher temperature isotherm, and heating  the \textit{p}\textsubscript{a} phase at 32~GPa followed by further compression. The results for the enthalpy, stoichiometry, and structural details of the phases obtained in these computational experiments are summarized in Fig.~2.

As evident from the data, compressing the gas mixture at higher temperature lowers the polymerization pressure and brings the polymer closer to CO$_2$ stoichiometry near 30~GPa. Oxygen undersaturation delays the onset of polymerization, while oversaturation does not accelerate it. On the other hand, oversaturation reduces the pressure at which oxygen is fully locked up in the polymeric structure. Thus, as a recipe for experimentalists to achieve a polymer with exact CO$_2$ stoichiometry, we propose a combination of oversaturation and compression at elevated temperatures starting from below 7~GPa. However, in the remainder of this paper, the analysis will focus on the properties derived from the \textit{p}\textsubscript{a} phase at 32~GPa and 300~K, corresponding to CO\textsubscript{1.88}+O$_2^{\mathrm{(\textit{s})}}$. Having a few non-reacted O$_2$ molecules, it provides a lower bound for the stability of the recovered polymer and its energy content.

Finally, we verify the mechanical stability of the recovered polymeric phase, which we denote as \textit{p}\textsubscript{a,rc}. The computed phonons do not exhibit any imaginary modes (see Supplementary Fig.~S5), which indicates that \textit{p}\textsubscript{a,rc} is metastable. However, it is important to note that this standard approach does not provide information about the magnitude of kinetic barriers and, consequently, the temperature range over which the system remains stable.

To further ascertain the kinetic stability of the recovered polymer, we performed two-phase simulations where 624-atom simulation cells were filled with 192 atoms of the gaseous \textit{m}\textsubscript{g}  phase and 432 atoms of the recovered polymeric  \textit{p}\textsubscript{a,rc} phase.
A \textit{NVT}-AIMD simulation was thereafter run for the two-phase mixture at  300~K for a total simulation time of 8.8~ps (the system is well-equilibrated by 3.1~ps), equilibrating to a near-ambient pressure of 1.6$\pm$0.6 kbar. As can be seen in Fig.~3a,b, there is coalescence of the polymeric chains with the gaseous mixture during the simulation. However, from the radial distribution function, $g(r)$, shown in Fig.~3c it is evident that the gaseous and amorphous parts of the mixture remain intact. The first peak of the $g(r)$, bifurcated  at 1.16~Å and 1.25~Å, corresponds to the original molecular gaseous phase and changes insignificantly, with a slight reduction in the second peak due to the coalescence described earlier. The small features between  1.3~Å and 2.8~Å, which are signatures of the polymeric structure, remain unchanged during the simulations.

To confirm the predictions of the AIMD simulations regarding the stability of the recovered polymer, we performed MLMD for a 15,600-atom two-phase cell over 100 ns. This system comprised 4,800 atoms in the molecular gas phase and 10,800 atoms in the amorphous \textit{p}\textsubscript{a,rc} structure. As shown in Figs. 3e and 3f, the two phases coexist without loss of kinetic stability for 100 ns, as indicated by the qualitatively unchanged radial distribution function. The mean square displacement (MSD) of different atom types provides a more intuitive perspective: C and O atoms initially part of the \textit{p}\textsubscript{a,rc} structure remain locked within it, while bulk gas and interfacial C and O atoms drift slowly until equilibrating at approximately 20 ns. The trapped \ce{O2}, as discussed earlier in Fig. 2, diffuses out of the amorphous structure without affecting the kinetic stability of the amorphous network.

The system equilibrates to a pressure of 1.9$\pm$0.3 kbar, indicating near-ambient, multi-nanosecond stability of the two-phase system. The validity of the MLMD simulations relies on the accuracy of the machine-learned interatomic potentials (MLIP), as demonstrated by the root mean square error (RMSE) in energies and forces shown in Fig. 3i. Additional validation was obtained by comparing the vibrational density of states (VDOS) across different timescales. The torsional and bond-bending modes, as well as their distributions, remain fundamentally unchanged from picosecond to nanosecond timescales, although some shuffling occurs among the layer and bond-stretch modes. This can be intuitively attributed to the release of trapped \ce{O2}, which induces more breathing vibrational modes within the system. As a result, any local layer-like characteristics are relaxed, allowing for increased lateral stretching in the amorphous network.

These results confirm the mechanical as well as kinetic stability of the polymeric phase at 300~K and a near-ambient pressure, with the highest temperature of stability being 700~K, when partial breakdown of the polymeric chain starts (discussed later).

\begin{figure*}[!hp]
\centering
\includegraphics[width=0.75\linewidth]{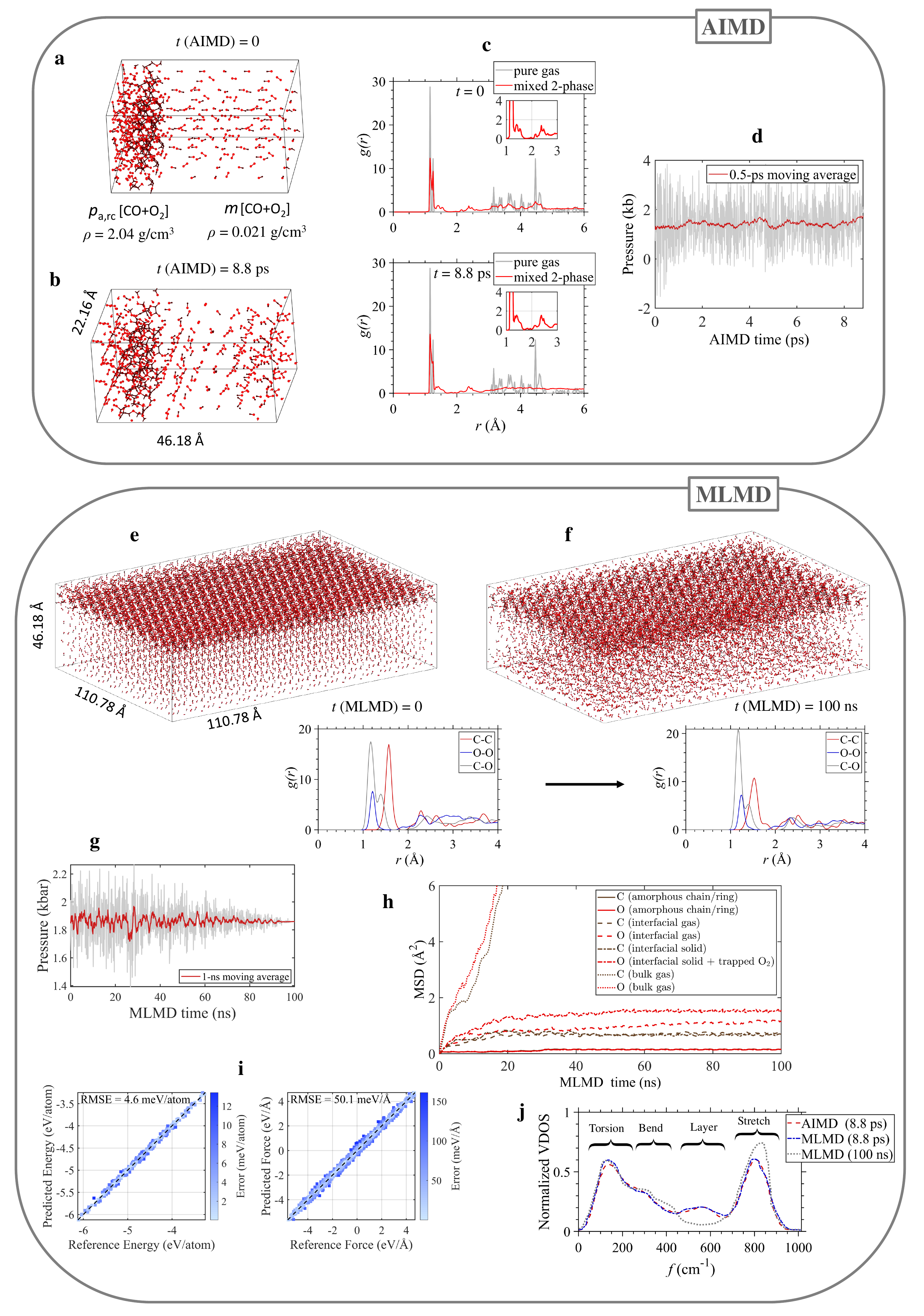}
\begin{center}
  \caption{Two-phase AIMD and MLMD results for the stability of polymeric amorphous material \textit{p}\textsubscript{a,rc}[CO+O\textsubscript{2}]. Visualization of the \textbf{a} initial and \textbf{b} final atomic configurations along a 624-atom two-phase simulation trajectory (432 atoms initially in \textit{p}\textsubscript{a,rc}[CO+O\textsubscript{2}] and 192 atoms in molecular $m$[CO+O\textsubscript{2}]
    at \textit{P} = 1.6$\pm$0.6 kbar and \textit{T} = 300~K. Oxygen and carbon atoms are shown in red and brown, respectively. \textbf{c} Radial distribution function, \textit{g}(\textit{r}), for the initial (top) and final (bottom) configurations. The latter is averaged over simulation time of 5.7 ps, after equilibrating for over 3.1 ps. The insets show the polymeric \textit{g}(\textit{r}) features that do not alter significantly during the course of the simulation. \textbf{d} Computed pressure during the AIMD simulation run is shown to remain at 1.6$\pm$0.6 kbar.
 \textbf{e,f} The 15,600-atom configurations at \textbf{e} the initial and \textbf{f} the final (after 100 ns) stages of an MLMD run, together with the associated radial distribution functions $g(r)$, demonstrating the stability of the recovered amorphous structure at larger length and time scales.
 \textbf{g} Computed pressure during the MLMD simulation, showning to equilibration at  1.9$\pm$0.3 kbar.
 \textbf{h} Mean square displacements of C and O atoms from the gaseous and polymeric phases during the MLMD run, demonstrating the stability of the latter.
 \textbf{i} Parity plots of energy and forces for the machine-learned interatomic potential (MLIP), with RMSE values indicated.
 \textbf{j}  Normalized vibrational density of states (VDOS) for AIMD (over 8.8 ps), MLMD (over the first 8.8 ps), and MLMD (over the entire 100 ns).}
\end{center}
\label{fig:recovery}
\end{figure*}


\subsection*{Stability comparison with respect to compression and recovery of initially molecular carbon dioxide}

Validating and understanding our theoretical findings for the polymerization of CO+O\textsubscript{2} requires an apt comparison with that of the well-studied CO\textsubscript{2} system. In this section, we examine comparatively the metastability of the two systems. For the latter, we consider three cases - crystalline CO\textsubscript{2}-V ($P2_12_12_1$)  and two amorphous solids generated as described below.

CO\textsubscript{2}-V, as alluded to earlier, is the only polymeric carbon dioxide phase which has been recovered to ambient conditions, but only at temperatures below 200~K\cite{yong2016crystal}. Because single-phase AIMD simulations usually cannot capture crystalline transitions, due to simulation cell size and time constraints, we have used the experimental  $P2_12_12_1$ structure as our starting point \cite{yoo1999crystal}. After equilibrating this structure at 60~GPa and 100~K, we decompressed it along the 100~K isotherm. However, at 4.9$\pm$0.2 kbar, when increasing the temperature to only 200~K, this crystalline phase breaks down to gaseous CO\textsubscript{2} (see Supplementary Fig.~S7) in complete agreement with the
experimental results. 

Next, we generated amorphous \textit{a}-CO\textsubscript{2} similarly to the way it was synthesized experimentally. For this purpose, 144-atom simulation cells with gaseous CO\textsubscript{2} were heated to high temperature at ambient pressure using a \textit{NVT} ensemble (to overcome kinetic reaction barrier) and subsequently compressed to pressures in the span of $\sim$60 to $\sim$118 GPa to overcome barriers for complete polymerization. After polymerization, they were slowly quenched to 300~K or below. For example, at $\sim$60 GPa, a kinetic barrier of roughly 600~K has to be overcome in order to obtain an amorphous structure on quenching to 300~K. This kinetic barrier is inversely proportional to the  pressure, and the barrier becomes only 300~K at $\sim$118 GPa. The resulting \textit{a}-CO\textsubscript{2} obtained this way  is a mixture of three- and four-coordinated carbon atoms. It can be recovered to ambient conditions only along the 200~K or lower temperature isotherm and breaks down to gaseous CO\textsubscript{2} when heated up (in single-phase \textit{NVT}-AIMD) to 300~K at 8.2$\pm$0.2 kbar (see Supplementary Fig.~S8). When decompressed along the 300~K isotherm, it breaks down to a molecular phase below $\sim$30 GPa.

Finally, for a direct comparison with the polymerized CO and O\textsubscript{2} mixture (referred to as \textit{p}\textsubscript{a,rc}[CO+O\textsubscript{2}] here and in the following section for clarity), we obtained polymeric \textit{p}-CO\textsubscript{2} (referred to as \textit{p}\textsubscript{a,rc}[CO\textsubscript{2}] in this section) directly by isothermal compression of 144-atom CO\textsubscript{2} simulation cells along the 300~K~isotherm. A polymeric phase emerges beyond $\sim$118~GPa comprised of entirely 4-coordinated carbon atoms. This is, in fact, the limiting case of $a$-CO\textsubscript{2}, as mentioned in the previous paragraph. As pressure is reduced,  \textit{p}\textsubscript{a,rc}[CO\textsubscript{2}] gradually changes from a fully four-coordinated to a mostly three-coordinated structure. Upon isothermal decompression to 6.2$\pm$0.2 kbar at 300~K, this polymeric phase persists but breaks down to gaseous CO\textsubscript{2} at temperature at or above 400~K in single-phase \textit{NVT}-AIMD simulations (see Supplementary Fig.~S9). 

Among the three aforementioned cases, only the latter yields a polymeric structure at ambient conditions. However, its stability is still significantly lower than that of \textit{p}\textsubscript{a,rc}[CO+O\textsubscript{2}], which persists at temperatures of at least 1000~K when heated at 7.7$\pm$0.3 kbar (see Supplementary Fig.~S9) as well as at $\sim$29 GPa (see Supplementary Fig.~S10) in single-phase simulations. 

However, single-phase (heat-until-breakdown) simulations may overestimate the stability of the simulated phase. To determine more accurately the temperatures at which the polymeric phases break down, we carried out two-phase \textit{NPT}  AIMD simulations. Here, the target pressure was set to 1 kbar instead of 1 atm, as barostat fluctuations at lower pressures can produce transient negative normal stress components. Under these conditions, it was observed that \textit{p}\textsubscript{a,rc}[CO\textsubscript{2}] breaks down to a gaseous molecular form at just 300~K and 3.2$\pm$0.1 kbar (see Supplementary Fig.~S11), while  \textit{p}\textsubscript{a,rc}[CO+O\textsubscript{2}] remains stable up to 700~K at 4.7$\pm$0.2 kbar (see Supplementary Fig.~S12 and S13), only beginning to partially break down at this temperature.

Additionally, we performed two-phase $NVT$  MLMD simulations using 11,520-atom cells composed of polymeric amorphous \textit{p}\textsubscript{a,rc}[CO+O\textsubscript{2}] and a gaseous phase (molecular \ce{CO2} or CO+\ce{O2}). As shown in Fig. 4, the structure remains intact at 500~K and only begins to break down upon heating to 700~K. This occurs regardless of whether the amorphous system is in contact with \ce{CO2} or CO+\ce{O2}.




Thus, our results predict that compressing CO+O\textsubscript{2}, rather than CO\textsubscript{2}, not only induces polymerization at lower pressures but also produces a polymer that is significantly more stable upon pressure release. Note that during AIMD simulations, negative (tensile) normal stress components along any axis can cause \textit{unphysical} fragmentation of the polymers. To avoid this, the simulations were equilibrated at a kilobar of pressure. However, the PBE exchange-correlation functional is known to overestimate the pressure in polymeric phases (as discussed in the 'Energetics' section). Therefore, these results can be used to ascertain the physics at ambient pressure.


\begin{figure*}[!ht]
\centering
\includegraphics[width=0.8\linewidth]{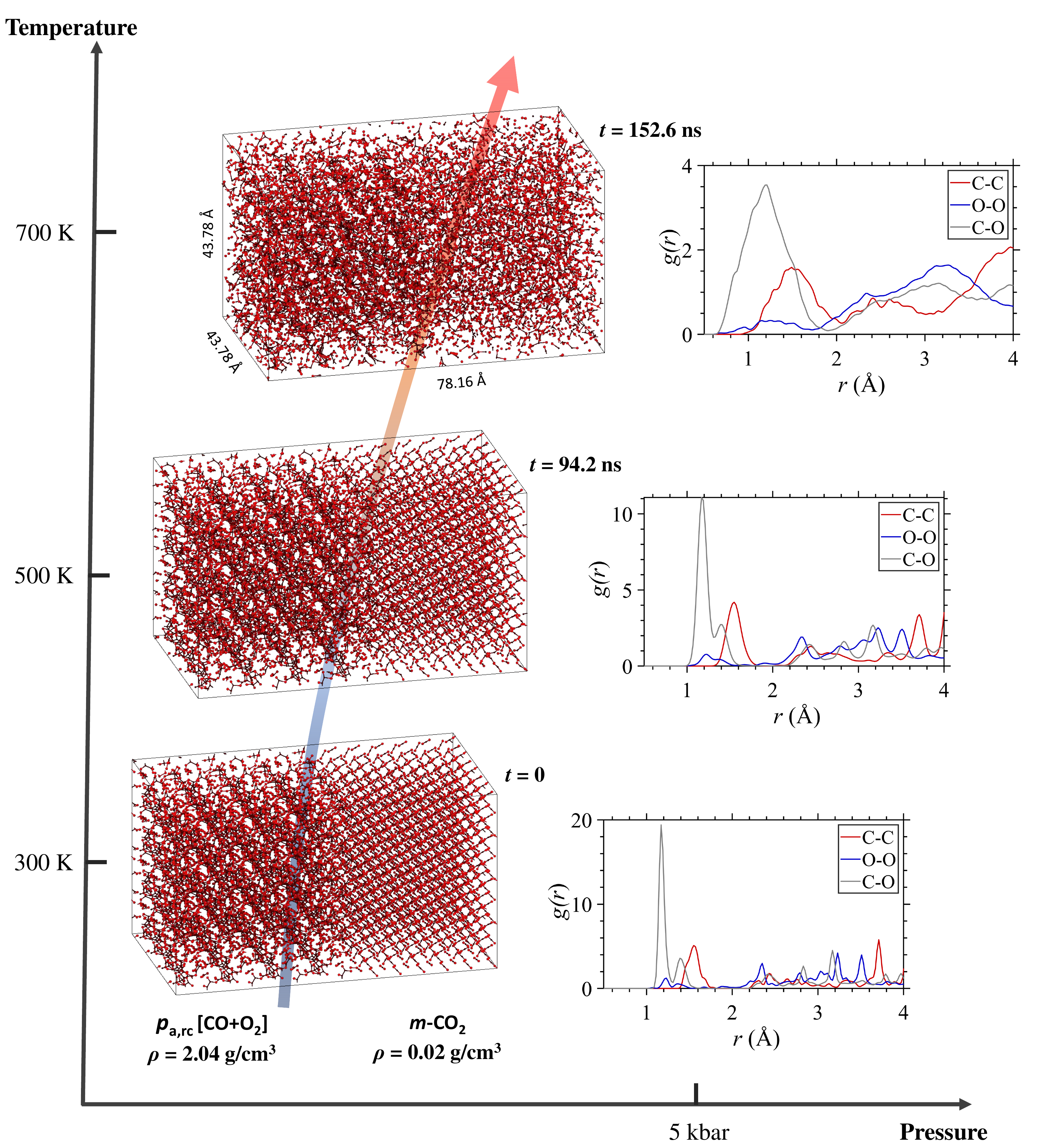}
\begin{center}
\vspace{1cm}
\caption{Kinetic stability of the polymeric amorphous \textit{p}\textsubscript{a,rc}[CO+O\textsubscript{2}] phase, assessed using 11,520-atom MLMD two-phase simulations (\textit{p}\textsubscript{a,rc}[CO+O\textsubscript{2}]  on one side and gaseous, molecular \textit{m}-\ce{CO2} on the other), by heating the two-phase cell to 500 and 700~K over a total simulation time of 152.6 ns. The polymeric amorphous phase begins to decompose at 700~K, as indicated by the topmost radial distribution function.}
\end{center}
\vspace{1cm}
\end{figure*}


\begin{figure*}[!ht]
\centering
\includegraphics[width=0.85\linewidth]{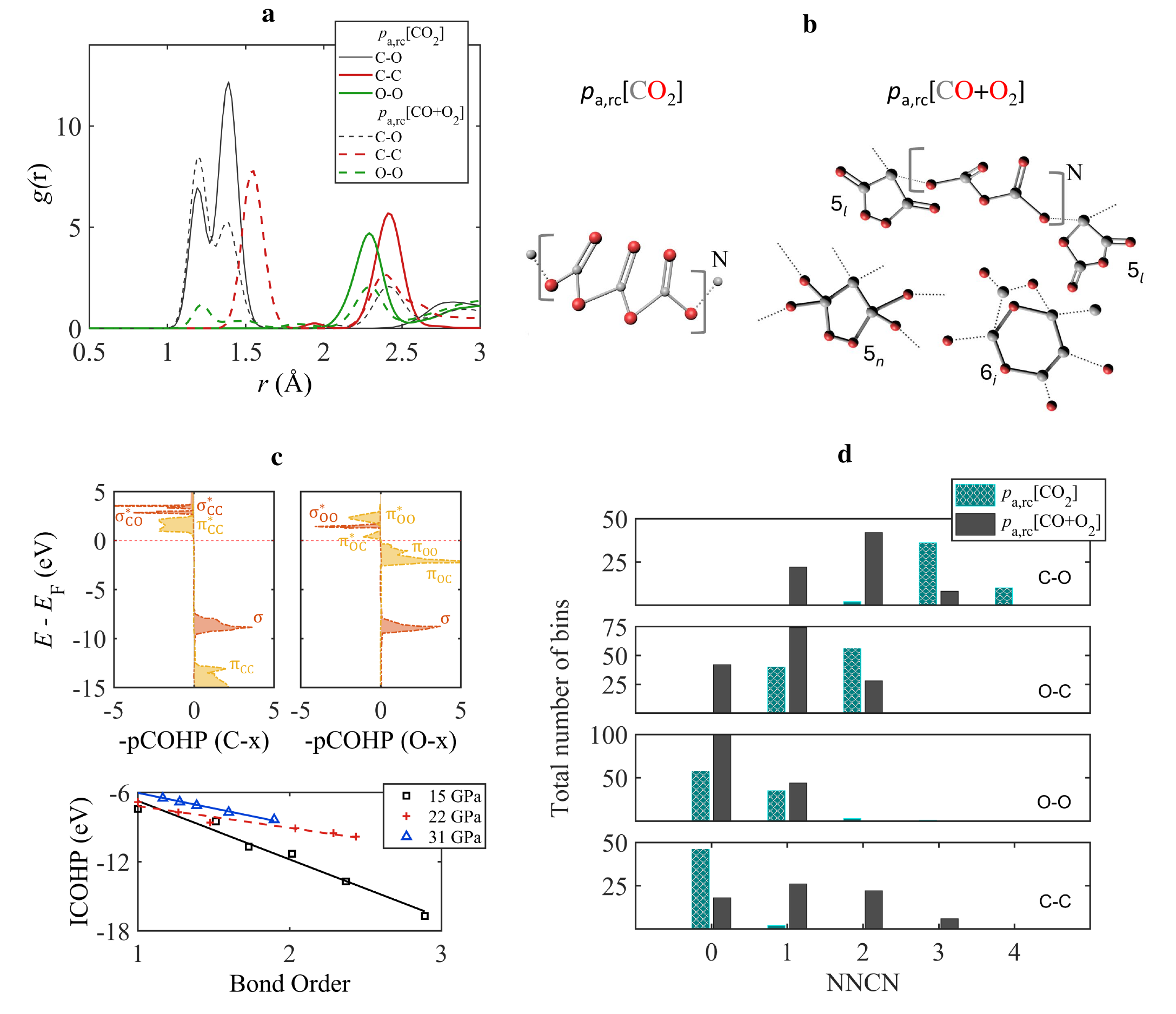}
\begin{center}
\vspace{1cm}
\caption{Comparative analysis of the structural features of the recovered polymeric phases of CO\textsubscript{2} (\textit{p}\textsubscript{a,rc}[CO\textsubscript{2}] at 200~K) and CO+O\textsubscript{2} (\textit{p}\textsubscript{a,rc}[CO+O\textsubscript{2}] at 300~K). \textbf{a} Pairwise radial distribution function, \textit{g}(r), for both systems at near-ambient pressures. \textbf{b} Comparison of chain structures, with carbon and oxygen atoms shown in grey and red, respectively. Here, $5_{\textit{l}}$, $5_{\textit{n}}$, and $6_{\textit{i}}$ indicate 5-member chain-linked, 5-member networked, and 6-member interlocked rings, respectively. \textbf{c} Partial crystal orbital Hamiltonian population (pCOHP) for \textit{p}\textsubscript{a}[CO+O\textsubscript{2}] for C-centered (top-left) and O-centered (top-right) bondings, as well as the integrated COHP as a function of bond order for selected pressures. \textbf{d} Nearest neighbor coordination number (NNCN) histograms for C$-$O, O$-$C, O$-$O, and C$-$C pairs.}
\end{center}
\vspace{1cm}
\end{figure*}

\subsection*{Structural properties}

The stability of polymers is directly linked to their atomic arrangements. To understand the differences among the recovered \textit{p}\textsubscript{a,rc}[CO+O\textsubscript{2}] and  \textit{p}\textsubscript{a,rc}[CO\textsubscript{2}] phases, we therefore present a comparative analysis of  their structural properties.  

A qualitative comparison of radial distribution functions in Fig.~5a shows that the \textit{p}\textsubscript{a,rc}[CO\textsubscript{2}] structure is only centered around C$-$O bonds. In contrast,  \textit{p}\textsubscript{a,rc}[CO+O\textsubscript{2}] has C$-$O, C=O, C$-$C, and O$-$O bonds, all contributing to the formation of chain-ring structures. Ball-and-stick models for the predicted chain or chain-ring structures are shown in Fig.~5b. The chain-ring structure in \textit{p}\textsubscript{a,rc}[CO+O\textsubscript{2}] is quite nuanced, with 5-member rings linked into the $-$O(CO)$-$ chains (shown as 5$_{\textit{l}}$), 5-member rings networked into a cage-type bonding mesh (shown as 5$_{\textit{n}}$), and 6-member rings interlocked with 5-member rings (shown as 6$_{\textit{i}}$). In contrast, the structure of \textit{p}\textsubscript{a,rc}[CO\textsubscript{2}] is just $-$O(CO)$-$ chains. 

The partial crystal order Hamiltonian population (pCOHP) \cite{dronskowski1993crystal,deringer2011crystal,maintz2016lobster} analysis of the \textit{p}\textsubscript{a}[CO+O\textsubscript{2}] phase, shown for C-centered and O-centered bonds in Fig.~5c, shows the usual C$-$O and O$-$O (single) $\sigma$-$\sigma$* bonding-antibonding orbital combinations and the O=C (double) $\pi$-$\pi$* combination. Of significance here are the unusually strong C=C $\pi$-$\pi$* bonds ($\Delta$\textit{E} $\sim$\ 16 eV) and C$-$C $\sigma$-$\sigma$* bonds ($\Delta$\textit{E} $\sim$\ 13 eV). The C=C $\pi$-$\pi$* bonds originate in the 6-member rings, which themselves are a result of including London dispersion forces in the simulations, while the C$-$C $\sigma$-$\sigma$* bonds are present in the 5-member rings. Recall that at ambient conditions, all stable molecules formed from elements in the second row of the periodic table have stronger $\sigma$ than $\pi$ bonds. The absence of these bonds in \textit{p}\textsubscript{a,rc}[CO\textsubscript{2}] (evident from the $g(r)$ in Fig.~5a) may be a contributing factor in its reduced kinetic stability during decompression.

Another important parameter when analyzing pressure-induced structures such as \textit{p}\textsubscript{a,rc}[CO+O\textsubscript{2}] is the bond order. For synthesizing energetic materials using a compression-decompression method, ideally, we would want a transition similar to the formation of \textit{cg}-N, where all triple bonds delocalize into single bonds. The integrated COHP (ICOHP) versus the bond order of \textit{p}\textsubscript{a,rc}[CO+O\textsubscript{2}] as a function of pressure is shown in Fig.~5c, the bond order being calculated using the integrated crystal order bond index (ICOBI) analysis \cite{müller2021crystal,maintz2016lobster}. As the pressure increases from 15 GPa to 31 GPa, the value of the \textit{maximum} bond order reduces from $\sim$3 to $\sim$2, indicating that the system is evolving towards a higher energetic state.

A distribution of the nearest-neighbor coordination numbers (NNCN) for the C$-$O, O$-$C, O$-$O, and C$-$C pairs in \textit{p}\textsubscript{a,rc}[CO+O\textsubscript{2}] and \textit{p}\textsubscript{a,rc}[CO\textsubscript{2}] is shown in Fig.~5d, with their mean values summarized in Table 1. The main observation here is the near absence of C$-$C and O$-$O bonding in \textit{p}\textsubscript{a,rc}[CO\textsubscript{2}]. The average C$-$O NNCN of $\sim$3, alongside O$-$C NNCNs of 1 and 2, also suggests the abundance of three-coordinated $-$O(CO)$-$ chains in that structure, which is in clear contrast to \textit{p}\textsubscript{a,rc}[CO+O\textsubscript{2}], where the chain intermittently has 5- and 6-member rings, thereby decreasing the mean C$-$O NNCN. The crystalline CO\textsubscript{2}-V ($P2_12_12_1$) phase, recovered to ambient pressure and 100 K, is also included in the table, again showing a complete absence of C$-$C bonding. It is important to note the existence of O$-$O pairs in both recovered systems, as mentioned earlier for the CO+O\textsubscript{2} case.

The structural differences among the various polymerized systems originate from their distinct reaction trajectories during compression, as discussed in the section on \textit{chemical dynamics} later. The linear geometry of O=C=O molecules imposes geometrical constraints on their approach to one another — constraints that are absent in the CO + O\textsubscript{2} mixture. These limitations inhibit C–C bond formation, resulting in differences in the polymerized structures that ultimately form.

In summary, the compression of a CO and O\textsubscript{2} mixture leads to the formation of a drastically different high-pressure structure compared to that of compressed CO\textsubscript{2}. The structural differences manifest in increased metastability of the polymerized amorphous product when it is recovered to ambient pressure.

\begin{table}[!ht]
\centering
\caption{Values of mean Nearest Neighbor Coordination Number (NNCN) calculated for 
different ion-type pairs in recovered phases of \textit{p}\textsubscript{a,rc}[CO\textsubscript{2}] at 200~K,  \textit{p}\textsubscript{a,rc}[CO+O\textsubscript{2}] at 300~K, and CO\textsubscript{2}-V at 100~K.}
\begin{tabular}{ >{\centering\arraybackslash}p{0.12\textwidth}  >{\centering\arraybackslash}p{0.07\textwidth}  >{\centering\arraybackslash}p{0.12\textwidth}  >{\centering\arraybackslash}p{0.07\textwidth} }
Mean NNCN& \textit{p}\textsubscript{a,rc}[CO\textsubscript{2}]& \textit{p}\textsubscript{a,rc}[CO+O\textsubscript{2}]& CO\textsubscript{2}-V\\
\midrule
Carbon-Oxygen& 3.08& 1.85& 3.93\\
Oxygen-Carbon& 1.78& 0.92& 1.95\\
Oxygen-Oxygen& \textbf{0.32}& 0.21& \textbf{0.00}
\\
Carbon-Carbon& \textbf{0.08}& 1.37& \textbf{0.00}
\\
\bottomrule
\end{tabular}
\end{table}

\subsection*{Energetics}

The potential use of \textit{p}\textsubscript{a,rc}[CO+O\textsubscript{2}] as an energetic material depends on the density of the recovered phase and its energy relative to the thermodynamically stable phase at ambient conditions, namely, molecular CO\textsubscript{2}. In this section, we focus on computing these properties. 

For a more detailed look at the energetics of the systems under consideration, we went beyond the PBE (GGA) exchange-correlation functional. In order to estimate the formation enthalpy of the final polymerized product more accurately, \textit{NVT}-AIMD simulations were performed for the full compression-decompression trajectories with PBE \cite{perdew1996generalized}, PBE + D3[BJ]  \cite{grimme2011effect}, PBE + D3[0] \cite{grimme2010consistent}, PBE + rVV10 \cite{peng2015scan+,roman2009efficient}, and SCAN-L \cite{mejia2017deorbitalization} + rVV10 combinations. Cell sizes with 432 and 864 atoms were used for the PBE and 258 atoms for the SCAN-L AIMDs (summarized in Supplementary Table~1). In addition, static single-point calculations were performed on five to seven configurations at each density along the PBE-calculated isothermal 108-atom AIMD trajectory using SCAN \cite{sun2012communication}, SCAN + rVV10 and HSE06 \cite{becke1992density} functionals, and then averaged. Fermi-Dirac smearing corresponding to a temperature of 300~K was used for these calculations. 

The pressure-density equations of state (EOS) for the CO+O\textsubscript{2} system are shown in Fig.~6a, calculated using the different combinations of exchange-correlation functions, van der Waals treatment and cell sizes (see Figs. S13 to S17 for a broader version).  The kinks in the EOS curves coincide with the initialization of $-$C$-$C$-$ polymerization. There is a general downward revision of pressure for the same density for SCAN-L and HSE06 w.r.t. PBE calculations, similar to observations of Bonev \textit{et al}. \cite{bonev2021energetics} on CO simulations for HSE06 w.r.t. PBE.  

The ranges of pressure over which $-$C$-$C$-$ and $-$C$-$O$-$ bonds and rings form are shown in Fig.~6b for the different combinations of cell size, exchange-correlation functional, and van der Waals treatment used. It can be seen that the polymerization process is delayed when using meta-GGA,  but the behavior of sequential bond formation remains consistent. With respect to the PBE + D3[0] 864-atom simulations, which form the bulwark of the analyses in this paper, the pressure domains for $-$C$-$C$-$, $-$C$-$O$-$ and rings are $\sim$7.1-8.5 GPa, $\sim$12.2-13.2 GPa, and $\sim$19.3-23.2 GPa, respectively.

Figure 6 shows the equations of state for the compression-decompression cycle of the CO+O\textsubscript{2} system from \textit{NVT}-AIMD simulations using an 864-atom cell, alongside the same cycle for 144-atom CO\textsubscript{2} cells at 200 K, all within PBE + D3[0]. The CO\textsubscript{2} system ends with an enthalpy higher than its starting molecular state, which is expected because molecular CO\textsubscript{2} is the most stable form at ambient pressure. A similar curve for the polymeric crystalline \textit{p}\textsubscript{c2}[CO+O\textsubscript{2}] system (obtained from metadynamics) was also evaluated, where it was decompressed to ambient conditions using \textit{NVT}-AIMD simulations in a 288-atom cell. However, this crystalline phase breaks down into the constituent CO and O\textsubscript{2} molecules below a pressure of $\sim$9 GPa and a density of $\sim$1.7 g/cc. This confirms that only the amorphous polymeric phase of CO+O\textsubscript{2} is recoverable.

\begin{figure}[!t]
\centering
\includegraphics[width=0.9\linewidth]{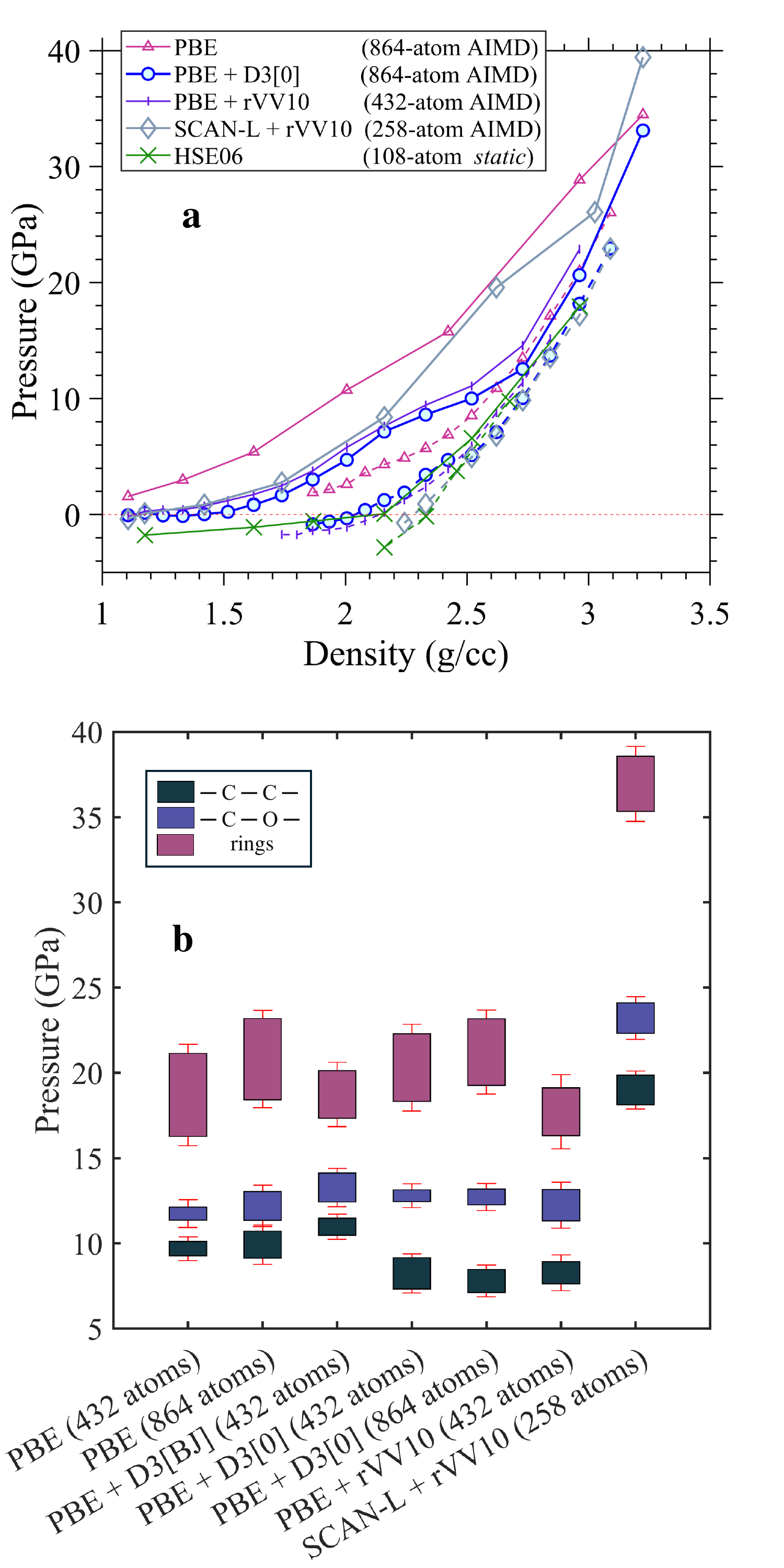}
\begin{center}
\vspace{1cm}
\caption{\textbf{a} Equations of state along the compression-decompression paths of CO+O\textsubscript{2} obtained within \textit{NVT}-AIMD at \textit{T}=300~K using various combinations of cell sizes, exchange-correlation functionals and van der Waals treatment. Solid and dashed lines indicate compression and decompression paths, respectively. \textbf{b} Pressure ranges over which the formation of $-$C$-$C$-$ and $-$C$-$O$-$ bonds and rings are observed under different simulation parameters. The red error bars are the standard deviations of the pressures in the ensemble-averaged \textit{NVT}-AIMD simulations.  
}
\end{center}
\vspace{1cm}
\end{figure}

The comparative analysis of density values, in Table 2, shows a compression of an initial gaseous mixture (\textit{m}\textsubscript{g}) with a density of 0.021 g/cc to a recovered polymer (\textit{p}\textsubscript{a,rc}) with density of 2.04 g/cc within PBE + D3[0]; this can be as high as 2.28 to 2.34 g/cc with corrections using meta-GGA or hybrid functionals. It is important to mention here that the deviation in the calculated values for the mass density and volumetric energy density is approximately 41\% and 32\%, respectively, although the gravimetric energy density varies only by 7\%.  The formation enthalpy of $\sim$204-243 kJ/mol ($\sim$4.6-5.5 kJ/g) is more than that of CO, which stands at $\sim$95 kJ/mol ($\sim$3.4 kJ/g) \cite{bonev2021energetics}, i.e., more stable. Furthermore, unlike CO, this composition is not susceptible to decomposition to graphite and oxygen, but converts to lower energy molecular CO\textsubscript{2} at a sufficiently high $T >$ 500~K. In essence, the process ends up yielding a denser polymer both gravimetrically and energetically. 

\begin{table*}[t!]
\centering
\caption{Interpolated values of enthalpy of formation ($\Delta_{f}$\textit{H}), density (\textit{$\rho$}\textsubscript{rc}) and volumetric energy density (\textit{E}\textsubscript{v}) for the decompressed/recovered polymeric CO+O\textsubscript{2 }phase \textit{p}\textsubscript{a,rc}, where $\Delta_{f}$\textit{H} has been calculated with respect to gaseous molecular CO\textsubscript{2} at ambient as detailed in Supplementary Tables S3, S4 and S5 of Supplementary Discussion 7}
\begin{tabular}{lllll}
\begin{tabular}[c]{@{}l@{}}Exchange-\\ Correlation used\end{tabular} & \begin{tabular}[c]{@{}l@{}}\textit{H}\textsubscript{\textit{p}\textsubscript{a,rc}} \\(eV/CO\textsubscript{2})\end{tabular} & \begin{tabular}[c]{@{}l@{}}$\Delta_{f}$\textit{H}
\\(kJ/mol)\end{tabular} & \begin{tabular}[c]{@{}l@{}}\textit{$\rho$}\textsubscript{rc} \\(g/cc)\end{tabular} & \begin{tabular}[c]{@{}l@{}}\textit{E}\textsubscript{v} \\(kJ/cc)\end{tabular}\\
\midrule

PBE (AIMD) & -20.707 & -222.173 & 1.615 & 8.153\\
PBE + D3[0] (AIMD) & -17.711 & -209.343 & 2.042 & 9.713 \\
SCAN-L + rVV10 (AIMD) & -24.109 & -207.484 & 2.281 & 10.754\\
SCAN + rVV10 (static) & -23.733 & -204.325 & 2.348 & 10.901\\
HSE06 (static) & -25.909 & -243.206 & 2.338 & 12.920\\
\bottomrule
\end{tabular}

\end{table*}

In terms of comparison to carbon dioxide, also shown in Fig.~7, the recovered \textit{p}\textsubscript{a,rc}[CO\textsubscript{2}] differs from the recovered \textit{p}\textsubscript{a,rc}[CO+O\textsubscript{2}] with a density higher by $\sim$0.1 g/cc and an enthalpy lower by $\sim$1 eV/CO\textsubscript{2} molecule. This suggests that for near-similar quantitative bulk properties of the recovered amorphous phase, a significant reduction in polymerization onset pressure has been achieved.

\section*{Phenomenological analysis of chemical dynamics}

To understand the formation of the p$_\mathrm{a}$ phase and its structural differences from the polymeric CO$_2$ phases, we analyze the atomic dynamics within the CO+O$_2$ and CO$_2$ systems before and after the onset of polymerization. We define the atomic number density, $\rho_N(r)$, as the number of atoms within a sphere of radius $r$ centered on an atom, averaged over all atoms and configurations in the AIMD. The sphere around each atom, inside which the first monomers start forming, has a radius equal to the first minimum of $\rho_N(r)$ and is called the kinetic sphere of radius $R_{\mathrm{ks}}$.

The evolution of the atomic arrangement in CO+O$_2$ along a compression trajectory --- from a nearly homogeneous molecular system at $t = 0$ (ambient pressure) to an aggregated one with $R_{\mathrm{ks}} = 5.3$~\AA\ at $t = 1$~ps ($P \approx 7$~GPa) --- is illustrated in Fig.~\ref{fig:8}b. Since $R_{\mathrm{ks}}$ is easily discernible only in the density region where the polymerization occurs, we define $R_{\mathrm{ks}}$ along the entire compression trajectory as:
\begin{equation}
R_{\mathrm{ks}}= \begin{cases}R_{\mathrm{ks}}\left(t^{\prime}\right) \times l(t) / l\left(t^{\prime}\right) & t \leq t^{\prime} \\ R_{\mathrm{ks}}(t) & t^{\prime \prime}>t>t^{\prime} \\ R_{\mathrm{ks}}\left(t^{\prime \prime}\right) \times l(t) / l\left(t^{\prime \prime}\right) & t \geq t^{\prime \prime}\end{cases}
\end{equation}

The mean distance between atoms $X$ (where $X$ is C or O) and all first-neighboring $Y$ atoms (C or O) inside their kinetic spheres, excluding the $Y$ atom in its own molecule, is the $X$--$Y$ kinetic diameter:
\begin{equation}
\bar{\phi}_1=\frac{1}{N_{\mathrm{X}}} \sum_{\mathrm{i}=1}^{N_{\mathrm{X}}} \sum_{\mathrm{j}=1}^{N_{\mathrm{Y}, \mathrm{ks}(\mathrm{i})}} \frac{\phi_{\mathrm{ij}}}{N_{\mathrm{ks}, \mathrm{Y}}(\mathrm{i})},
\end{equation}

where $N_{\mathrm{X}}$ is the total number of X atoms in the cell and $N_{\mathrm{Y}, \mathrm{ks}}(\mathrm{i})$ is the total number of first-neighbor Y atoms inside the kinetic sphere of the $\mathrm{i}^{\text {th }} \mathrm{X}$ atom, but not bonded with it. During this computation, we ensure that the following condition is satisfied:

\begin{equation}
\sum_{\mathrm{j}=1}^{N_{\mathrm{O}, \mathrm{ks}}(i)} 2 \pi\left(1-\cos \Omega_{\mathrm{j}}\right) \lesssim 4 \pi
\end{equation}

where $\Omega_{\mathrm{j}}\left[=\cos ^{-1}\left(\varphi_{\mathrm{j}} / r_{\text {shell }}\right)\right]$ is the subtended solid angle of the $\mathrm{j}^{\text {th }} \mathrm{Y}$ atom with respect to a shell of radius $r_{\text {shell }}$. Here the radial distance between the $\mathrm{n}^{\text {th }}$ and $\mathrm{n}+1^{\text {th }}$ shells calculated radially outwards from the X atoms, $\Delta r_{\text {shell }}\left[=r_{\text {shell, } \mathrm{n}+1}-r_{\text {shell, } \mathrm{n}}\right]$, is the tunable parameter used to ensure that the total subtended solid angle is approximately equal to $4 \pi \mathrm{sr}$.

\begin{figure}
\centering
\includegraphics[width=0.75\linewidth]{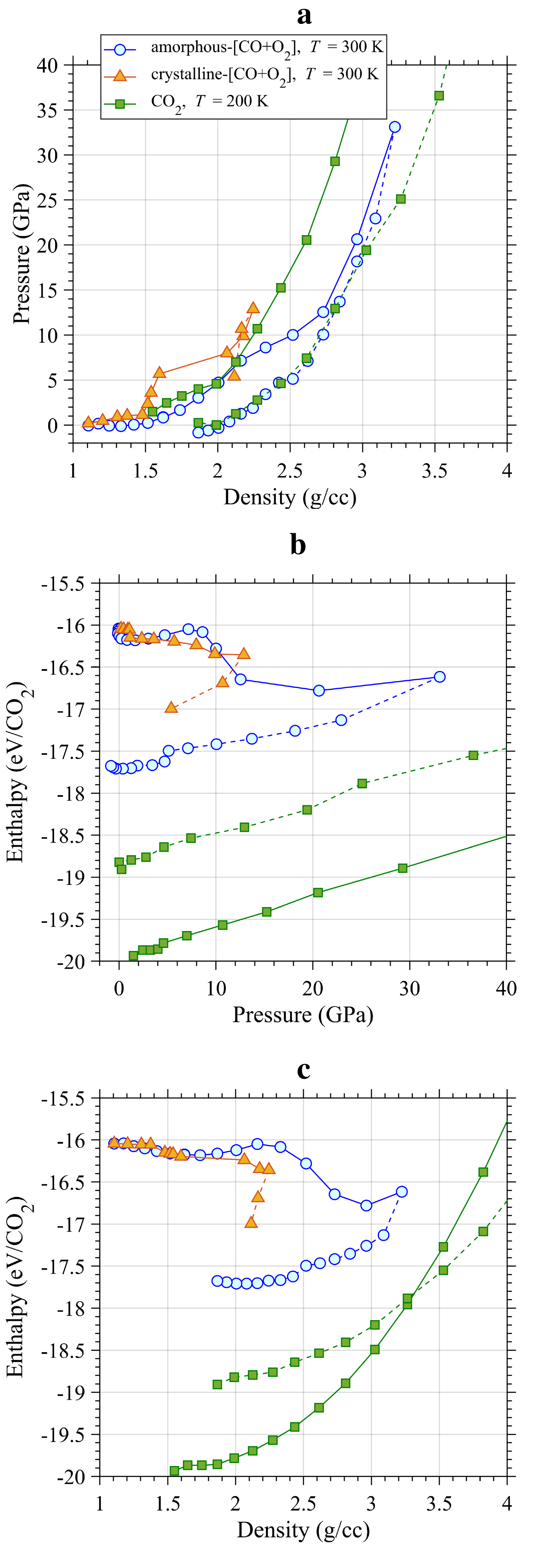}
\begin{center}
\caption{Equations of state along the compression-decompression paths of the amorphous \textit{p}\textsubscript{a,rc}[CO+O\textsubscript{2}] obtained using \textit{NVT}-AIMD at \textit{T}=300 K, the polymeric crystalline \textit{p}\textsubscript{c2}[CO+O\textsubscript{2}] obtained with metadynamics along the compression path and subsequently decompressed using \textit{NVT}-AIMD at \textit{T}=300 K, and the compound CO\textsubscript{2} obtained with \textit{NVT}-AIMD at \textit{T}=200 K are presented in \textbf{a} pressure-density, \textbf{b} enthalpy-pressure, and \textbf{c} enthalpy-density spaces. The solid and dashed lines represent the compression and decompression paths, respectively. All curves are evaluated within PBE + D3[0], using 432-atom, 288-atom, and 144-atom simulation cells for amorphous CO+O\textsubscript{2}, crystalline CO+O\textsubscript{2}, and CO\textsubscript{2}, respectively. The curves corresponding to CO+O\textsubscript{2} in Fig.~6a are the same as the PBE + D3[0] data shown in Fig.~1b. The data for CO\textsubscript{2} is only shown up to 40 GPa, while the polymerization of this system commences beyond $\sim$118 GPa at 300 K.  
}
\end{center}
\end{figure}

\begin{figure*}[!ht]
\centering
\includegraphics[width=\linewidth]{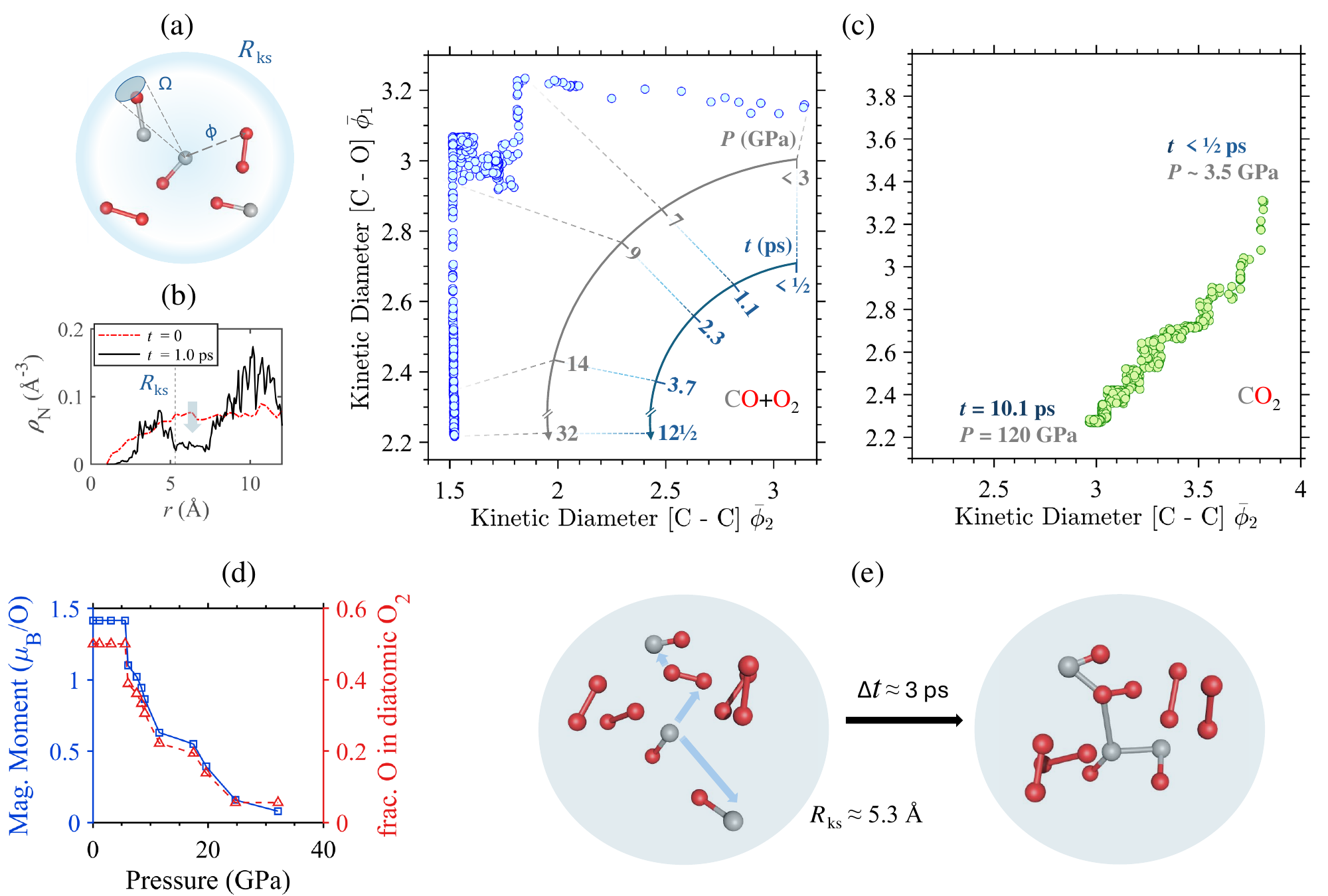}
\caption{a An illustration of kinetic sphere of radius $R_{\mathrm{ks}}$ (shaded as a blue circle) and kinetic diameter $\phi$ defined in the text. b Radial number density, $\rho_{\mathrm{N}}$, for the $\mathrm{CO}+\mathrm{O}_2$ system at ambient pressure ( $t=0$ ) and the onset of polymerization ( $t=1.0 \mathrm{ps}$ ) along a compression trajectory. c Evolution of the kinetic $[\mathrm{C}-\mathrm{O}] \bar{\phi}_1$ versus $[\mathrm{C}-\mathrm{C}] \bar{\phi}_2$ diameters along isothermal compression simulations starting from ambient pressure at 300 K . The $\mathrm{CO}+\mathrm{O}_2$ and $\mathrm{CO}_2$ systems are compressed up to 32 and 120 GPa , respectively. The two radial axes for $\mathrm{CO}+\mathrm{O}_2$ show the time and pressure for specific points of significance along the AIMD trajectory. d Change in average magnetic moment per oxygen atom and fraction of residual molecular $\mathrm{O}_2$ as a function of pressure. e Atomic arrangements from an AIMD trajectory showing the relative arrangements of CO and $\mathrm{O}_2$ molecules in the vicinity of $\mathrm{O}_4$ clusters, together forming the kinetic sphere.}
\label{fig:8}
\end{figure*}

The evolution of the C--O versus C--C kinetic diameters along an 864-atom AIMD compression trajectory spanning pressures from ambient to $\sim 32$~GPa during a 12.5~ps simulation is shown in Fig.~\ref{fig:8}c. Here $t'$ and $t''$ are estimated at 0.5~ps and 7.6~ps, respectively, for the CO+O$_2$ system. The resulting locus of points exhibits a $\Gamma$-shaped path. Note that a similar observation holds true even if just the nearest or the average of the first two to five neighboring atoms is used to define the diameter (see Supplementary Fig.~S19).

The C--C polymerization happens just beyond $\sim 7$~GPa, which corresponds to the corner of this $\Gamma$-shaped path, where the average C--C diameter is minimized; presumably the system overcomes the activation barrier for polymerization. Thereafter, as the C--C polymerization completes by $\sim 9$~GPa, relaxation of the C--O diameter is observed, most likely an effect of C--O polymerization which completes by $\sim 14$~GPa. The cumulative simulation time needed for these two stages of polymerization is approximately 3~ps, which gives a benchmark for the time scales of bond rearrangement in such compressed carbon--oxygen systems. This is comparable to the experimentally observed timescales of 1--3~ps in both ionic as well as cyclic covalent compounds.\textsuperscript{1--3}

In the case of CO$_2$, a $\Gamma$-shaped path is not followed by the kinetic diameters, despite a simulation time of 10~ps, even though the system eventually reaches a four-coordinated amorphous structure. This highlights the fact that, although the CO+O$_2$ mixture and CO$_2$ are stoichiometrically equivalent, the dynamics of their reaction trajectories are completely different. The linear O{=}C{=}O molecules impose geometrical constraints on their approach to one another that are not present in the CO+O$_2$ mixture, thereby inhibiting C--C bond formation.

Another observation is the local reaction environment inside the simulation cell. In a 2:1 CO:O$_2$ mixture, the residual magnetic moment of oxygen in the cell is shown in Fig.~\ref{fig:8}d as a function of pressure. The average magnetic moment is $1.41~\mu_B$ at ambient and decreases with increasing pressure. This is consistent with an ample amount of free O$_2$ molecules in the CO+O$_2$ cell at lower pressures, including conditions where C--C and C--O bonds form. To visualize the effect of the residual O$_2$, we look inside a kinetic sphere during a simulation. As shown in Fig.~\ref{fig:8}e, O$_2$ molecules tend to aggregate into O$_4$ clusters, which is similar to the $\varepsilon$ phase of oxygen at comparable pressures.\textsuperscript{4,5} They provide a local potential energy environment within which CO and other O$_2$ molecules react to form C--C and C--O bonds. The elimination of such an environment inhibits polymerization. This shows that even though the mixture is stoichiometrically equivalent to the compound CO$_2$, the local chemical environment during polymerization is completely non-stoichiometric.


\section*{Discussion}

Density functional theory-based methods were used to predict a new, low-pressure way to synthesize a metastable, polymeric amorphous form of \ce{CO2} that can be recovered at ambient conditions and used as a high-energy-density material. Traditional methods for making fully 4-coordinated polymeric \ce{CO2} require extreme pressures (approximately 118 GPa), and the resulting phases revert to molecular \ce{CO2} when pressure is released at room temperature. In contrast, we propose compressing a mixture of CO and \ce{O2} (with a stoichiometry equivalent to \ce{CO2}) along the 300K isotherm. This approach lowers the polymerization onset to about 7~GPa and achieves complete C–C polymerization by around 23~GPa, which is a significant reduction in the required pressure. When decompressed back to ambient pressure at 300~K, this polymeric phase remains stable with almost no change in bonding statistics, showing both mechanical and kinetic stability.

Constrained metadynamics searches identified two transient crystalline intermediates that are slightly favorable below 10~GPa but quickly become amorphous when compressed further. This confirms that the amorphous phase is the main product at higher pressures. Small amounts of unreacted \ce{O2} remain up to about 30~GPa, and later form oxygen chains that become part of the polymer without disrupting its structure. This effect increases the kinetic barriers and therefore enhances the metastability of the material.

Kinetic stability was further tested using two-phase AIMD simulations (624 atoms) for more than 8~ps, and large-scale MLMD simulations (15,600 atoms) for 100~ns. Both showed that the polymer and gas phases can coexist, with stable pressures (1.6–1.9 kbar) at room temperature and locked-in C,O frameworks. Unreacted \ce{O2} was able to diffuse out without causing the polymer to collapse. Phonon analysis showed no imaginary modes, confirming mechanical metastability. Both MLMD and two-phase $NPT$ AIMD simulations also demonstrated that the amorphous polymer structure remains stable up to 700~K at ambient pressure before it begins to partially break down. A comparative analysis with the well-studied \ce{CO2} system shows that our approach successfully reproduces the experimentally observed stability of known polymeric-[\ce{CO2}] phases, thereby validating our methodology.

Structural comparisons between recovered polymeric-[CO+\ce{O2}] and conventional polymeric-[\ce{CO2}] reveal key differences. Polymeric-[CO+\ce{O2}] exhibits C–O, C=O, C–C, and O–O bonds forming complex chain-ring networks (including 5- and 6-membered rings), whereas polymeric-[\ce{CO2}] consists mainly of three-coordinated −O(CO)− chains with minimal C–C or O–O bonding. Crystal Orbital Hamilton Population (COHP) analyses highlight unusually strong C=C $\pi$-$\pi^*$ and C–C $\sigma$-$\sigma^*$ interactions in polymeric-[CO+\ce{O2}], indicating higher stored energy. Energetic metrics predict higher formation enthalpies (–204 to –243~kJ/mol), densities (2.04–2.34~g/cc), and volumetric energy densities (8.2–12.9 kJ/cc)—surpassing polymeric \ce{CO2} and rivaling TNT energetics. 


If the prediction that compression of an initial metastable mixture lowers the onset of polymerization and yields kinetically robust HED materials is generally true, then this opens up opportunities to create and recover various energetic materials starting from initial mixtures of C,H,O,N compounds as the starting building blocks. To this end, we have provided Raman signatures of the relevant polymeric and amorphous phases in Supplementary Discussion 10 for aiding in experimentation.

\section*{Methods}
\subsection*{\textit{Ab initio} one-phase simulations}
\textit{Ab initio} molecular dynamics (AIMD) simulations, utilizing the Born–Oppenheimer approximation \cite{born1985quantentheorie}, with temperature being controlled via a Nosé–Hoover thermostat in a canonical (constant-\textit{NVT}) or an isothermal–isobaric (constant-\textit{NPT}) ensemble, form the backbone of the methods used in this paper. 

The Vienna \textit{Ab Initio} Simulation Package (VASP) \cite{kresse1993ab,kresse1996efficiency,kresse1996efficient} was used for such calculations, where 4- and 6-electron \textit{hard} projector augmented wave (PAW) pseudopotentials were used for carbon and oxygen respectively, with a 850 eV plane-wave cutoff energy for generalized gradient approximation (GGA)\cite{perdew1992atoms} and 1000 eV for meta-GGA exchange-correlation functionals.  Most of these calculations were performed using the $\Gamma$-point for sampling (resolution of at least $< 2\pi \times$ 0.10~Å$^{-1}$) the first Brillouin zone (1BZ) but for some cases, we used a single special \textbf{k}-point (1/4, 1/4, 1/4), as was introduced by A. Baldereschi \cite{baldereschi1973mean}.  

Supercells were used, with 108, 192, 216, 258, 432 and 864 atoms, with an ionic time step of 0.5-0.8 fs for AIMDs at ambient pressure and temperature. Cubic supercells were used to prevent a bias towards any non-amorphous structure. For equilibrating simulations along these compression and decompression paths, the combination of ionic step size and total number of steps was such that the total physical time for the simulation system was always more than 5 ps, exceeding 10 ps in pivotal cases (see Supplementary Table~1 and 2 for complete list). 

Oxygen was considered to be magnetic in all production simulations (see Supplementary Table~1) by utilizing spin-polarized calculations. This allows the converged ground-state solution to be the triplet (total spin \it{S} \rm= 1) state, rather than the singlet state \cite{freiman2015magnetic}. 

The PBE (Perdew-Burke-Ernzerhof) formulation \cite{perdew1996generalized} of the GGA exchange-correlation functional was mostly used for the simulations. However, to check for the effects of climbing up Jacob's ladder, as the sensitivity of reaction kinetic barriers to the nature of exchange-correlation is of utmost importance \cite{kaplan2023understanding,kanungo2024unconventional}, the SCAN (strongly constrained and appropriately normed) \cite{sun2012communication} and SCAN-L \cite{mejia2017deorbitalization} meta-GGA functionals were also used. The hybrid functional HSE06 (Heyd–Scuseria–Ernzerhof) \cite{becke1992density} was also used for static tests. For van der Waals dispersion to be taken into account, the DFT-D3 [zero damping] \cite{grimme2010consistent}, DFT-D3 [Becke-Johnson damping] \cite{grimme2011effect}, and the rVV10 (revised Vydrov–van Voorhis) non-local correlation functional \cite{peng2015scan+,roman2009efficient} were used. Thus, the combinations of PBE, PBE + D3[BJ], PBE + D3[0], PBE + rVV10, and SCAN-L +rVV10 were used for AIMD simulations, while PBE, SCAN, SCAN + rVV10 and HSE06 were used for single-point calculations at multiple snapshots along the compression and decompression AIMD trajectories. 

The compression and decompression analysis of any system using \textit{NVT}-AIMD simulations, starting with a gaseous mixture configuration at $<$0.1 kbar, involved a two-step approach. First, a fast compression trajectory was simulated using short simulations of 500 ionic steps at each volume along an isotherm, starting from ambient conditions. The volume was decreased by 1-2\% at each point on the trajectory. Upon reaching the furthest point along the isotherm where complete pressure-induced polymerization is observed, equilibration simulations were performed for 5-10 ps, as described earlier (see Supplementary Table~1), at this furthest point as well as at intermediate points along the compression path for finding the resulting equilibrated ionic arrangement and constructing the equation of state. Second, for recovery, a fast decompression trajectory was similarly simulated with the volume being increased by 1-2\% at each point on the trajectory. Equilibrating AIMD simulations of 5-10 ps were also performed at select points on the decompression trajectory to find the resulting equilibrated ionic arrangement and constructing the recovery equation of state. 

Single-phase MLMD simulations were also performed using LAMMPS \cite{thompson2022lammps} for $\sim$ns timescale checks, and those can be found in Supplementary Discussion S3. Extensive discussion on Allegro \cite{tan2025high, musaelian2023learning, kozinsky2023scaling} training is also available in Supplementary Discussion S9.

\subsection*{\textit{Ab initio} two-phase simulations}
For the two-phase simulations, we performed a few cases with \textit{NVT} ensemble while the rest were \textit{NPT} ensemble simulations. In each of these cases, we started with cuboid supercells, with each side's phase being well-equilibrated beforehand with respect to atomic arrangements using \textit{NVT} simulations. The two-phase simulations used Baldereschi \textbf{k}-point for sampling, with 192 atoms for \textit{m}-CO\textsubscript{2} $\vert$ \textit{p}\textsubscript{a,rc}[CO\textsubscript{2}], and 216 to 624 atoms (see Supplementary Table~2) for \textit{m}-[CO+O\textsubscript{2}] $\vert$ \textit{p}\textsubscript{a,rc}[CO+O\textsubscript{2}] systems respectively. Here, the symbol $\vert$ is used to denote each of the two phases on either side. Then a AIMD simulation was performed on these merged cells with a time step of 0.5~fs, with a total simulation time exceeding 5~ps for all cases.

For the larger-scale 15,600-atom \textit{m}-[CO+O\textsubscript{2}] $\vert$ \textit{p}\textsubscript{a,rc}[CO+O\textsubscript{2}] and 11,520-atom \textit{m}-CO\textsubscript{2} $\vert$ \textit{p}\textsubscript{a,rc}[CO+\ce{O2}] two-phase \textit{NVT} MLMD simulations, machine learned interatomic potentials (MLIPs) trained using Allegro \cite{tan2025high, musaelian2023learning, kozinsky2023scaling} was used to perform molecular dynamics simulations using LAMMPS \cite{thompson2022lammps}. The training was performed within the RMSE threshold of 5 meV/atom for energies and 50 meV/Å for forces.

\subsection*{Metadynamics}
Constrained \textit{NPT}-metadynamics calculations were performed, with the PBE + D3[0] exchange-correlation functional and the bias potential being constructed of fixed Gaussians, at each \textit{$\rho$},\textit{T} point after constraining the C$-$O and O$-$O bond lengths at higher pressures to within 2.5\% of ambient values (as observed in AIMD at corresponding pressures) and using the cell basis vectors (magnitude as well as angles) as collective variables. Cell sizes corresponding to 36- to 108-atoms were used for the simulations, in increments of 6 atoms (i.e. 2CO+O\textsubscript{2}). It is worth mentioning that, as a check, metadynamics simulations were also performed without constraining bond lengths, which yielded molecular crystal CO\textsubscript{2}-I as the stable solution. This is unphysical given the fact that at 300~K, the system's thermal energy is insufficient for the reaction of CO and O\textsubscript{2}, even at 30 GPa. 

\subsection*{Vibrational spectra}
Phonon and Raman spectra calculations were performed, in 108-atom cells with PBE + D3[0] functional, utilizing density functional perturbation theory (DFPT) \cite{baroni2001phonons} calculations for the Born effective charge (BEC) tensor and Phonopy \cite{togo2023first} was utilized for evaluating the dynamical response of the system. Raman spectra was calculated (see Supplementary Discussion 10 for formulation) using this BEC tensor. 

\hfill \break
Input, configuration, and trajectory files for compression and decompression pathways have been uploaded at \url{https://doi.org/10.5281/zenodo.15027031} for the sake of repeatability.


\section*{Acknowledgements}

We thank C.S. Yoo for discussions. This work was performed under the auspices of the U.S. Department of Energy by Lawrence Livermore National Laboratory (LLNL) under contract number DEAC52-07NA27344. The authors acknowledge funding support from the DOE Laboratory Directed Research and Development (LDRD) program at LLNL under the project tracking code 23-ER-028.

\section*{Author contributions}

J.C.C. and S.A.B. designed research; R.P. performed research; S.A.B. supervised research; R.P. and S.A.B. wrote the paper; J.C.C. provided guidance; all authors discussed the results and contributed to the final manuscript.

\end{document}